\documentclass[%
 reprint,
 amsmath,amssymb,
 aps,
twocolumn
]{revtex4-2}

\usepackage{graphicx}
\usepackage{dcolumn}
\usepackage{bm}
\usepackage{hyperref}

\newcommand{\bra}[1]{\left\langle #1\right|}
\newcommand{\ket}[1]{\left|#1\right\rangle}
\newcommand{\abs}[1]{\left|#1\right|}

\begin{document}

\title{High performance imaging of $^{171}$Yb atoms in shallow clock-magic tweezers by alternating dual-tone narrowline cooling}

\author{Yunheung Song$^{1}$}
\email{ysong@kriss.re.kr}
\author{Kangheun Kim$^{1}$}
\author{Jeong Ho Han$^{1}$}
\author{Seungtaek Oh$^{1,2}$}
\author{Jongchul Mun$^{1}$}
\email{jcmun@kriss.re.kr}
\affiliation{%
$^{1}$ Korea Research Institute of Standards and Science (KRISS), Daejeon 34113, Republic of Korea\\
$^{2}$ Graduate School of Quantum Science And Technology, KAIST, Daejeon 34141, Republic of Korea
}%
\date{\today}

\begin{abstract}
We demonstrate imaging $^{171}$Yb single atoms in clock-magic tweezers of 759.4 nm wavelength, with above 99.9\% fidelity and survival. We use alternating dual-tone narrowline imaging for more efficient three-dimensional cooling in tweezers, allowing several-millisecond imaging in 200 $\rm\mu K$ trap depth, which is half of typical depth used for imaging in clock-magic tweezers. Accordingly, even without repumping, imaging survival is still close to 99.9\% with the high fidelity, which can enable high performance nondestructive qubit measurements based on metastable shelving. Moreover, our simulation predicts that more optimal configuration could further reduce the trap depth, as improving the imaging performance. This imaging capability in shallow traps opens high performance imaging for more general trap wavelength, and lays the foundation for large scale systems over 1,000 qubits and highly repeatable tweezer clocks.
\end{abstract}

\maketitle

\section{Introduction}

Single atom array in optical tweezers armed with scalability and programmability in both static and dynamic configurations has emerged as one of the most promising modalities for quantum information science. Among various atomic elements, alkaline-earth-like atom (AEA) array~\cite{Kaufman2018,Endres2018,Thompson2019} is getting attention as an emerging frontier thanks to their rich energy structure, allowing for various new quantum control capabilities in quantum computing~\cite{AtomComputing2024,Endres2024,Thompson2025_0}, metrology~\cite{Kaufman2024,Endres2024_2,Kaufman2020}, and networking~\cite{Goban2025,Covey2025}. Especially, fermionic AEAs have Zeeman manifolds of only nuclear spin in $J=0$ ground and metastable states, which can be used as a highly coherent qubit~\cite{AtomComputing2022,Kaufman2022,Thompson2022}, and affords optical clock transitions without high B field to mix triplet states~\cite{Clockreview}. Among them, $^{171}$Yb is a unique stable isotope of nuclear spin 1/2, which can naturally encode a qubit, and takes numerous quantum control advantages in mid-circuit operations~\cite{Kaufman2023,AtomComputing2023,Covey2023}, multi-qubit encoding~\cite{Covey2024}, quantum error correction~\cite{Thompson2022_2,Thompson2025_0}, etc.

Meanwhile, extending system size larger and larger is gaining much interest~\cite{Birkl2024,AtomComputing2024_2,Zeiher2024,Pan2025,Endres2025,Lukin2025}. High performance imaging of atom array is important not only for precise qubit measurement and frequency metrology~\cite{Kaufman2019,Endres2019}, but also for achieving large system size of no defect~\cite{Pan2025,Endres2025}, regardless of continuous loading development~\cite{AtomComputing2024_2,Zeiher2024,Lukin2025,Thompson2025}. In this regard, imaging atoms above 99.9\% fidelity and survival is an important milestone for a large system over 1,000 qubits.

Although high performance imaging above 99.9\% fidelity and survival has been demonstrated for AEAs~\cite{Endres2019_2,AtomComputing2025}, that level of high performance imaging is nontrivial in general, especially for fermionic AEAs because of two main reasons. First, due to the energy structure of AEAs, scattering by trap light during imaging is connected to several loss channels such as metastable and ionized states. The previous high performance imaging was demonstrated with efficient repumping of trappable scattered states~\cite{Endres2019_2,AtomComputing2025}, which is not possible for general trap wavelengths, including clock-magic traps of $^{171}$Yb required for many essential applications~\cite{AtomComputing2024,Kaufman2024,Endres2024_2,Kaufman2020,Goban2025,Covey2025,Kaufman2019,Endres2019,Kaufman2023,Covey2024}. Thus, the ability to reduce trap depth during imaging, thereby minimizing trap scattering, is a key to the high performance, as well as an advantage of more trap numbers~\cite{Schreck2022}. This is also important for nondestructive qubit and clock measurements based on metastable shelving~\cite{Kaufman2019,Endres2019,Kaufman2023} where repumping is limited. Second, the nuclear spin degeneracy of fermionic AEAs generates a dark state which prevents seamless laser cooling. To tackle this problem, previous methods require zero B field~\cite{Thompson2022}, spatial polarization gradient~\cite{Kaufman2023,Covey2023}, and stretched transition in high B field~\cite{AtomComputing2023,Covey2023}. However, none of these methods have demonstrated imaging capability in sufficiently shallow trap depth for the high performance.

\begin{figure}
\includegraphics[width=0.48\textwidth]{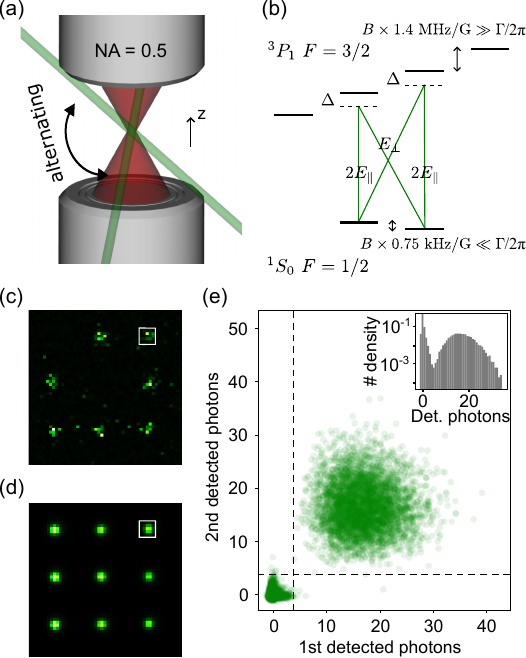}
\caption{(a) Schematic of alternating cooling in a tweezer trap. (b) Dual-tone driving of ${^1S_0},F=1/2-{^3P_1},F=3/2$ transitions of $^{171}$Yb atom is shown with relative strength. With a moderate B field (15 G for our case) and the natural linewidth $\Gamma=2\pi\times182$ kHz, the ground state splitting is negligible, while the excited state can be selectively driven by each tone. (c,d) $3\times3$ $^{171}$Yb array image (c) of a single shot and (d) averaged over 20,000 shots by 5.4 ms alternating dual-tone narrowline imaging. (e) Correlation plot between detected photons of two sequential shots in the $5\times5$ pixels region of interest (ROI) in (c) and (d). Dashed lines indicate optimized thresholds. Inset exhibits log-scale histogram for the total detected photons in the ROI.}
\label{fig_concept}
\end{figure}

Here, we demonstrate imaging $^{171}$Yb array above 99.9\% fidelity and survival for the first time in clock-magic tweezers, to our knowledge. Even without repumping, the survival is still about 99.9\% thanks to imaging capability in shallow traps. To avoid a dark state with the nuclear spin degeneracy, we use dual-tone frequencies~\cite{Thompson2025,KRISS2024} to simultaneously address $\ket{{^1S_0},F=1/2,m_F=\pm1/2}-\ket{{^3P_1},F=3/2,m_F=\pm1/2}$ narrowline transitions in moderate B field (Fig.~\ref{fig_concept}(b)), and this scheme can be described in terms of Morris-Shore (MS) transformation~\cite{MS1983}. Also, we suggest alternating 2-axis imaging for more efficient 3D cooling in tweezers (Fig.~\ref{fig_concept}(a)), and the experimental performance is compared with a Monte Carlo (MC) simulation for laser cooling dynamics. Finally, the alternating dual-tone narrowline cooling allows imaging $3\times3$ $^{171}$Yb array in 200 $\rm\mu K$ trap depth, resulting in above 99.9\% fidelity and survival (Fig.~\ref{fig_concept}(c-e)) with comparable imaging time of several milliseconds.

\section{Dual-tone narrowline driving}
\label{sec_dualtone}
\begin{figure}
\includegraphics[width=0.48\textwidth]{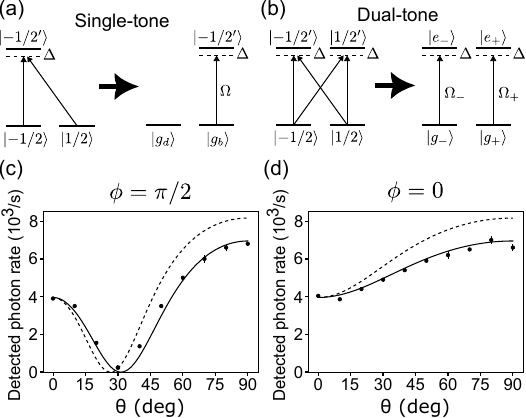}
\caption{(a,b) Narrowline driving of ${^1S_0},F=1/2-{^3P_1},F=3/2,\abs{m_F}=1/2$ transition. (a) Single-tone driving results in a lambda system, which is MS-transformed to a two-state system and a dark state. (b) Dual-tone driving results in a double lambda system, which is MS-transformed to 2 separate two-state systems. (b,c) Dependence of detected photon rate for dual-tone driving on polarization angle $\theta$ with phase difference (b) $\phi=\pi/2$ and (c) $\phi=0$, respectively, of which electric field is $\vec E=e^{i\phi}E\cos\theta\hat{e}_\perp+E\sin\theta\hat{e}_\parallel$. Solid (dashed) lines show theoretical values based on Eq.~\ref{eq_scattering} with (without) considering transmission difference between P and S polarizations at glass cell. For these measurements, fixed input imaging beam intensity ($s=2$ for $\theta=0^\circ$) and detuning ($\Delta=-0.92\Gamma$) are used. Error bar indicates standard error of the mean.}
\label{fig_MStransform}
\end{figure}

Consider a narrow intercombination line of $^{171}$Yb, $^1S_0,F=1/2-{^3P_1},F=3/2$ transition shown in Fig.~\ref{fig_concept}(b). With a moderate B field (15 G for our setup), ground states are nearly degenerate due to insensitivity of nuclear spin on B field, and each excited state can be separately driven thanks to narrow transition linewidth of an order of $\Gamma=2\pi\times182$ kHz. In clock-magic tweezer of 759.4 nm wavelength, $^1S_0,F=1/2-{^3P_1},F=3/2,\abs{m_F}=1/2$ transitions have the smallest differential AC Stark shift and the most efficiently collectible dipole radiation pattern among the transitions~\cite{Thompson2022,Kaufman2023,Covey2023}, so we choose the transitions for more homogeneous and efficient narrowline imaging across an atom array.

When a single excited state $\ket{-1/2'}\equiv\ket{^3P_1,F=3/2,m_F=-1/2}$ is selected by a single frequency of detuning $\Delta$ (Fig.~\ref{fig_MStransform}(a)), the system of interest is a lambda system that consists of $\ket{\pm1/2}\equiv\ket{^1S_0,F=1/2,m_F=\pm1/2}$ and $\ket{-1/2'}$. Because a lambda system can always be transformed to a two-state system $\ket{g_b}-\ket{1/2'}$ and a dark state $\ket{g_d}$~\cite{GAFbook}, driven ground states are optically pumped to the dark state and no photon is emitted at steady state.

On the other hand, when both $\ket{\pm1/2'}\equiv\ket{^3P_1,F=3/2,m_F=\pm1/2}$ are chosen by a dual-tone driving of the same intensity and detuning $\Delta$ for each transition~\cite{Thompson2025,KRISS2024} (Fig.~\ref{fig_MStransform}(b)), the system of interest is a double lambda system of $\ket{\pm1/2}$ and $\ket{\pm1/2'}$. For this system of degenerate manifolds, MS transformation~\cite{MS1983} can be used to express it as 2 separate two-state systems, $\ket{g_-}-\ket{e_-}$ and $\ket{g_+}-\ket{e_+}$, of Rabi frequencies $\Omega_-$ and $\Omega_+$, respectively. For the sake of simplicity, we suppose the k vector of the driving laser beam and the quantization axis (B field) are perpendicular. Then, the Rabi frequencies are given by
\begin{align}
\Omega_{\pm}\propto\sqrt{\abs{E_{\perp}}^2+4\abs{E_{\parallel}}^2\pm2\abs{E_{\perp}}\abs{E_{\parallel}}\sqrt{2(1-\cos2\phi)}},
\label{eq_MSRabi}
\end{align}
where $\phi$ is the phase difference between $E_{\perp}$ and $E_{\parallel}$ (see Appendix~\ref{sec_MS} for more detail and general case). For the MS-transformed two-state systems, steady-state solution~(see Appendix~\ref{sec_steady}) gives total scattering rate $R_{\rm tot}$ as
\begin{align}
R_{\rm tot}=\frac{2R_ + R_-}{R_+ + R_-}=\frac{\Gamma}{2}\frac{s_{\rm tot}}{1+s_{\rm tot}+(2\Delta/\Gamma)^2},
\label{eq_scattering}
\end{align}
where $R_\pm$ is two-level scattering rate for $\ket{g_\pm}-\ket{e_\pm}$ transition, and $s_{\rm tot}=2s_+s_-/(s_+ + s_-)$ is total intensity saturation parameter with that of each transition,  $s_\pm=2\Omega_\pm^2/\Gamma^2$. The scattering rate has the same form as that of a simple two-state system, so conventional Doppler cooling theory~\cite{lasercooling} can be applied to explain the system.

There are two special cases, $\phi=\pi/2$ and 0, where difference between $\Omega_+$ and $\Omega_-$ is maximized and minimized, respectively. At first, we consider $\phi=\pi/2$ case. Now, Eq.~\ref{eq_MSRabi} is reduced to $\Omega_{\pm}\propto\abs{E_{\perp}}\pm2\abs{E_{\parallel}}$, and this predicts $\Omega_-=0$ at $\abs{E_{\perp}}=2\abs{E_{\parallel}}$, i.e. $\ket{g_-}$ becomes a dark state. To verify this, we measure collected photons during 10 ms exposure of an imaging beam~(see Appendix~\ref{sec_exp} for more details), as a function of polarization angle $\theta$, which determines ratio between electric field components as $\abs{E_\parallel/E_\perp}=\tan\theta$. The polarization angle is adjusted by a half wave plate followed by a fixed quarter wave plate to set phase difference $\pi/2$ between the two E field components. The result in Fig.~\ref{fig_MStransform}(c) agrees well with the theoretical prediction, and nearly no photon is emitted around $\theta=30^\circ$ due to the dark state. 

Second, in $\phi=0$ case which is guaranteed if an imaging beam is linearly polarized regardless of its k vector, the Rabi frequencies always have the same magnitude, $\Omega_\pm=\Omega_{\rm tot}\propto\sqrt{\abs{E_{\perp}}^2+4\abs{E_{\parallel}}^2}$, so the total scattering rate in Eq.~\ref{eq_scattering} is described like a two-state system of intensity saturation parameter $s_{\rm tot}=2\Omega_{\rm tot}^2/\Gamma^2$, and no dark state occurs in any case. We measure similarly the scattering rate as shown in Fig.~\ref{fig_MStransform}(d), and it shows good agreement with the prediction. In the remaining works in this paper, we use linear polarization, so $\phi=0$ case for dual-tone imaging beams.

\section{Alternating cooling in tweezer}

\begin{figure}
\includegraphics[width=0.48\textwidth]{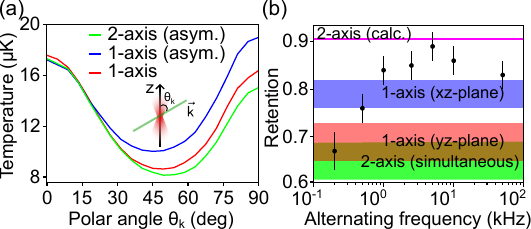}
\caption{(a) MC simulation of Doppler cooling in a tweezer trap as a function of polar angle $\theta_k$ between k-vector direction of imaging beam and tweezer axis, with intensity $s=2$ and detuning $\Delta=-\Gamma$. 1-axis cooling in radially symmetric trap, 1-axis and 2-axis cooling in asymmetric trap (10\% anisotropy) are compared. (b) Release and recapture measurement for atom temperature after 2-axis alternating cooling depending on alternating frequency, with $s=2$, $\Delta=-\Gamma$, and release time of 20$~\rm\mu s$. Colored shades indicate the measurement results without alternating in various configurations. Purple solid line is MC simulation for 2-axis cooling.}
\label{fig_cooling}
\end{figure}

Although we show that dual-tone narrowline driving of $^{171}$Yb atom allows a description of light scattering like a simple two-state system, exact behavior in a tweezer trap is not described by conventional Doppler cooling theory based on atoms in free space~\cite{lasercooling} where 3D cooling requires 3 axes of laser beams. In a tweezer trap, where tight radial directions have degenerate trap frequencies and axial direction has an order of magnitude lower trap frequency, 1-axis cooling by a single oblique retro-reflected beam which covers both radial and axial directions~\cite{IW1982} can result in 3D cooling thanks to trap anharmonicity~\cite{Regal2012,anharmonic2020}. In Fig.~\ref{fig_cooling}(a), we check this by classical MC simulation considering motion in Gaussian beam potential, momentum damping and diffusion from steady-state laser scattering (see Appendix~\ref{sec_MC}). Atom temperature in trap is calculated as a function of the polar angle $\theta_k$ between tweezer axial direction and k vector of an imaging beam. In the calculation, z axis is the tweezer axis, and y axis is the quantization axis. For 1-axis cooling with a retro-reflected imaging beam in xz plane (red solid line), the k vector which covers only axial ($\theta_k=0$) or radial ($\theta_k=90^\circ$) direction results in relatively high temperature where loss is too high at low trap depth (see Appendix~\ref{sec_MC}). While, if the k vector covers both axial and radial direction, more efficient 3D cooling occurs as expected, and the minimum temperature is achieved around $\theta_k=50^\circ$. However, in the real world, tweezer traps are not perfectly symmetric along radial directions, and the anisotropy in trap waists is commonly 10\% level~\cite{Sylvain2024}. With this anisotropy, the simulation shows higher temperature due to less efficient anharmonic mixing (blue solid line). Thus, utilizing 2-axis cooling with 2 retro-reflected imaging beams in xz and yz planes (green solid line), we can achieve even lower temperature than 1-axis cooling in symmetric trap, despite the anisotropy.

We experimentally address this issue by measuring atom temperature by release and recapture method~\cite{Grangier2008}. After laser cooling of fixed total intensity applied on atom ($s=2$) and detuning ($\Delta=-\Gamma$) for 20 ms with various other setting, atom is released from the trap for 20~$\rm \mu$s and recaptured, and then the retention is measured as shown in Fig.~\ref{fig_cooling}(b). Unexpectedly, either the retention of 1-axis cooling with an imaging beam in xz plane ($\theta_k=52^\circ$, blue shade) or with an imaging beam in yz plane ($\theta_k=73^\circ$, red shade) is higher than that of 2-axis cooling by simultaneously using the two imaging beams (green shade), though calculation shows higher retention (purple solid line). We attribute this to spatial polarization gradient which contains both linear and circular polarization components formed by interference between 2 axes. To resolve this problem, we alternate 2 axes of imaging beams and the results as a function of alternating frequency are shown in Fig.~\ref{fig_cooling}(b). At low alternating frequency, an atom reaches thermal equilibrium of 1-axis cooling before switching to the other axis, and the resultant temperature is similar to that of 1-axis cooling. At moderate alternating frequency of kHz range, atom temperature approaches closely to the value from the MC simulation for 2-axis cooling.

\begin{figure}
\includegraphics[width=0.48\textwidth]{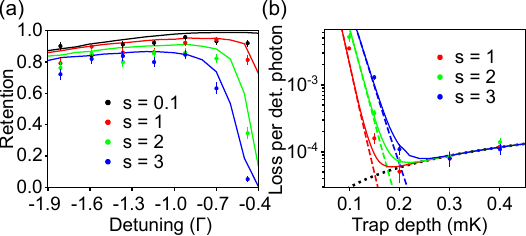}
\caption{(a) Temperature measurement by 20-$\rm\mu s$ release and recapture for alternating dual-tone imaging. Colored dots are measured data with specified intensity saturation parameters, and solid lines are corresponding MC simulation results. (b) Atom loss per detected photon during the imaging. Colored dots are experimental data with $s=1,2,3$ and $\Delta=-0.77\Gamma,-\Gamma,-1.1\Gamma$, respectively. Red, blue, and green dashed lines indicate corresponding MC simulations, and black dotted line is calculated trap Raman scattering to metastable states. Colored solid lines are summation of the losses from the MC simulation and the Raman scattering. For (a) and (b), alternating frequency is 2.5 kHz.}
\label{fig_scan}
\end{figure}

In Fig.~\ref{fig_scan}, we scan control parameters for the alternating cooling and compare the results with theory. In Fig.~\ref{fig_scan}(a), atom temperature is measured similarly as in Fig.~\ref{fig_cooling}(b), as varying detuning at a fixed intensity. The minimum atom temperature appears at a detuning which is 10 - 20\% larger magnitude than that from conventional theory~\cite{lasercooling}, in a good agreement with the MC simulation which predicts about 4.9 $\rm\mu K$, 7.1 $\rm\mu K$, 8.8 $\rm\mu K$, and 10.3 $\rm\mu K$ for $s = 0.1$, 1, 2, and 3, respectively. Our simulation observe that the optimal detuning becomes close to the theoretical value $-\sqrt{1+s}~\Gamma/2$ when trap is magic for the cooling transition.

In Fig.~\ref{fig_scan}(b), atom loss per detected photon during alternating cooling is measured by retention after applying 310 ms of imaging beams, as a function of trap depth at a fixed intensity. At low trap depth ($<0.2$ mK), atom loss is dominated by momentum damping \& diffusion dynamics (red, green, and blue dashed lines), but, at high trap depth ($>0.2$ mK), trap Raman scattering to metastable states from $^3P_1$ (black dotted line) is dominant~(see Appendix~\ref{sec_scattering}). Note that the Raman scattering rate to strongly anti-trapped $^3P_2$ is more than 4 times higher than that to $^3P_0$. Around 0.2 mK trap depth, atom loss is minimized, and the lower intensity, the lower atom loss in our parameters where the loss from trap lifetime is not significant.

\section{High performance imaging demonstration}

\begin{figure}
\includegraphics[width=0.48\textwidth]{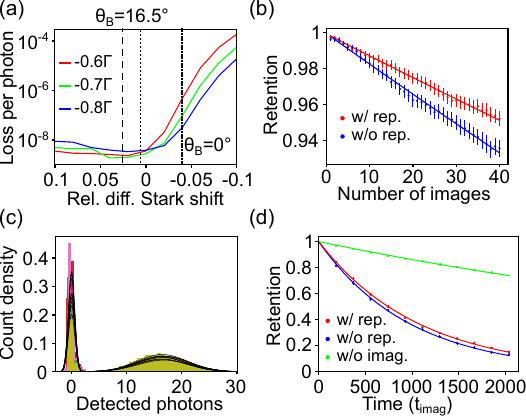}
\caption{High fidelity and survival imaging of $^{171}$Yb array of 0.2 mK trap depth with intensity $s=1$, detuning $\Delta=-0.77\Gamma$, imaging time $t_{\rm imag}=5.4~{\rm ms}$, and alternating frequency of 5 kHz. (a) Calculation of loss per emitted photon as a function of relative differential AC Stark shift of the imaging excited state. Dashed and dotted lines indicate the Stark shifts of ${^3P_1},F=3/2,m_F=\pm1/2$ states, respectively, when the angle $\theta_B$ between B field (15 G) and trap E field (depth 0.2 mK) is about $16.5^\circ$. Dash-dotted line corresponds to the Stark shift at $\theta_B=0^\circ$. (b) Atom retention after repeated imagings, with or without repumping beams. (c) Histogram of detected photons for each site fitted by Gaussian plus exponential distributions (black solid lines). (d) Imaging lifetime measurement with or without repumping beams. For comparison, data with only trap beam is also plotted.}
\label{fig_image}
\end{figure}

For high performance imaging of $3\times3$ ytterbium 171 array, we utilize the alternating dual-tone narrowline cooling with parameters shown in the caption of Fig.~\ref{fig_image}. We set the angle $\theta_B$ between the B field and the trap polarization as $16.5^\circ$, in order to make the differential AC Stark shift of the excited states smaller and red-detuned for less loss (Fig.~\ref{fig_image}(a)), with which the dominant loss source is trap Raman scattering plus a little bit of vacuum loss during imaging. Note that there is no perfect imaging-magic angle~\cite{Kaufman2018,Kaufman2023} for our B field (15 G) due to higher order contributions. In addition, we apply 1389 nm and 770 nm repumping beams during imaging to save some population pumped into metastable states by the Raman scattering, via ${^3P_0},F=1/2-{^3D_1},F=3/2$ and ${^3P_2},F=2\pm1/2-{^3S_1},F=3/2$ transitions, respectively (see Appendix~\ref{sec_exp} for more details). For the latter transition, Zeeman and trap tensor shifts make the frequency range broad, but we only use single tone for each F level with experimentally optimized detuning of -85 MHz from free-space resonance. After loading single atoms, we repeat an experimental sequence to take consecutive 41 shots of the $3\times3$ array image in the same parameters with exposure time of 5.4 ms and fixed gap time of 20 ms. A single shot image is shown in Fig.~\ref{fig_concept}(c), and Fig.~\ref{fig_concept}(d) is the averaged image over 20,000 shots.

To analyze imaging performance, we use $5\times5$ pixels regions of interest (ROI) like an example ROI shown in Fig.~\ref{fig_concept}(c,d). For the ROI, Fig.~\ref{fig_concept}(e) represents a correlation plot between photons of two sequential shots, and the inset exhibits a corresponding log-scale histogram of detected photons. For more quantitative characterization, we firstly fit the retention from 41 consecutive images in Fig.~\ref{fig_image}(b) and it gives imaging survival of 99.949(5)\% averaged across the array, after accounting for the loss from 20 ms gap time with trap lifetime of 28(1) s. Also, we fit histograms of detected photons by Gaussian and exponential distributions~\cite{Endres2018} as shown in Fig.~\ref{fig_image}(c), resulting in 99.935(8)\% average imaging fidelity. Second, to get imaging fidelity and survival in a complementary way, we apply a model-free method~\cite{Kaufman2018,Scazza2025} to the repeated two sequential shots of 20 ms gap time. It gives average imaging fidelity of 99.957(6)\% and survival of 99.94(3)\% after accounting for the gap time loss. In Fig.~\ref{fig_image}(d), lifetime with imaging beams continuously on is measured in unit of the imaging time 5.4 ms. The fitted imaging lifetime is 1,080(60) shots and corresponding 1-shot survival probability $\sim99.91\%$ is a bit smaller than the value measured with consecutive images.

For comparison, similar experiments are also performed without repumping beams. By the fitting method, average imaging fidelity and survival (gap-time-corrected) are 99.935(8)\% and 99.902(5)\%, respectively. On the other hand, by the model-free method, average imaging fidelity and survival (gap-time-corrected) are 99.94(1)\% and 99.92(3)\%, respectively. Lastly, the measured imaging lifetime is 980(40) shots.

\section{Discussion}

First, we discuss a measurement fidelity for a qubit state based on our imaging scheme. According to a state-of-the-art qubit detection in clock-magic tweezer to utilize clock shelving~\cite{Kaufman2023}, we estimate a ground nuclear-spin qubit detection fidelity, assuming numerical aperture upgrade to 0.6 (reference~\cite{Kaufman2023}) from 0.5 (ours) and the same clock-transition fidelity. Compared with the reference, our trap depth is half, while imaging time 3.66 ms (for maintaining photon numbers of the current 5.4 ms imaging) is similar. Therefore, trap Raman scattering, which is a dominant source for error and loss, becomes half, resulting in about 99.4\% and 99.2\% estimated detection fidelity for imaged and shelved qubit states, respectively. Assuming perfect clock transitions, these increase to about 99.9\% and 99.5\% with loss less than 0.1\% and 0.3\%, respectively, which corresponds to ground-clock detection performance.

Second, we discuss possible further improvement. Trap Raman scattering is still a primary loss mechanism even with repumping. Our simulation predicts that making our imaging polar angle $\theta_k=52^\circ,73^\circ$ (limited by current geometry) close to the optimal would allow to further reduce trap depth to $\sim140~{\rm\mu K}$ as improving imaging performance~(see Appendix~\ref{sec_MC}). In addition, multi-tone repumping beams to cover the broad frequency range of ${^3P_2},F=2\pm1/2-{^3S_1},F=3/2$ transitions could help improve the performance.

Finally, we outline possible methods for local imaging. Although our scheme requires two radial directions with considerable tilt angle for axial component, appropriate beam shaping to homogeneously and exclusively illuminate an imaging zone within the intersection between the two imaging beams (see Fig.~\ref{fig_concept}(a)) could be used for zone-based architecture~\cite{Lukin2023}. Also, using local metastable shelving~\cite{Kaufman2023} or local hiding light~\cite{AtomComputing2023} will allow local imaging with global addressing.

\section{Conclusion}

In summary, we have demonstrated above 99.9\% fidelity and survival for imaging $^{171}$Yb array in shallow clock-magic tweezers by alternating dual-tone narrowline cooling, within several-millisecond timescale. Our method brings high performance imaging capability for more general trap wavelength, while demanding less repumping. This will play a crucial role in large scale systems over 1,000 qubits and highly repeatable tweezer clocks. We believe that our alternating cooling approach and its analysis can be similarly applied to other imaging schemes for not just Yb but also other atomic elements for more efficient 3D cooling in tweezers, without interference between axes.

\begin{acknowledgements}
This work was supported by the National Research Foundation of Korea (NRF) under Grant No. RS-2023-NR119928, RS-2023-00283259, and RS-2025-25464182.
\end{acknowledgements}

\begin{appendix}
\section{Experimental details}
\label{sec_exp}
In our experimental apparatus, ytterbium 171 atoms from a homemade oven kept in 420$^\circ$C are firstly cooled and trapped by 2D MOT of two retro-reflected 399 nm laser (Toptica DL Pro HP) beams of 40 mW power and 4 mm beam radius, detuned -60 MHz from ${^1S_0},F=1/2-{^1P_1},F=3/2$ resonance in about 70 G/cm B-field gradient formed by permanent magnets. Then, the atoms are pushed to a rectangular glass cell (Akatsuki Technology) by a 399 nm beam, and captured by 3D MOT of singlet transition (blue MOT) in 27 G/cm B-field gradient, which has higher capture velocity than triplet transition. The 399 nm laser (Toptica DL Pro HP) beams for blue MOT have 3 mm beam radius, 20 mW power per beam, and detuning -45 MHz from the resonance. After 500 ms loading of blue MOT, the 2D MOT beams and the push beam are turned off and 556 nm beams for 3D MOT of triplet transition (green MOT) are turned on, while switching B-field gradient to 9 G/cm. The 556 nm laser beams have 3 mm beam radius, 12 mW power per beam, -8 MHz detuning from ${^1S_0},F=1/2-{^3P_1},F=3/2$ resonance, of which frequency is modulated by amplitude of 7.8 MHz and frequency of 150 kHz. Next, the green MOT is compressed by continuously reducing beam power to 75 $\rm\mu W$ per beam and detuning to -0.2 MHz for 150 ms. The 556 nm laser is generated by a second harmonic generation (SHG) module (NTT WH-0556-000-A-B-V) from 1112 nm tapered amplifier (TA, Toptica TA Pro). All the 399 nm lasers are frequency-stabilized by offset frequency locking to a reference laser frequency-locked to a ULE cavity. The 556 nm laser is frequency-locked directly to the ULE cavity.

After the compressed MOT stage, we use clock-magic tweezers of 759.356 nm wavelength for atom loading. The laser for the tweezers is a homemade external cavity diode laser (ECDL) frequency-locked to a wavemeter (HighFinesse Angstrom WS/8 IR) and amplified by a homemade TA, of which ampified spontaneous emission~\cite{Ludlow2021} is filtered by a volume Bragg grating of 50 GHz bandwidth. After fiber and an acousto-optic modulator (AOM) for switching, maximum available power is about 400 mW, and the laser is transferred to a spatial light modulator (SLM, Meadowlark Optics UHSPDM1K-759-PC8) after beam expansion. Then, the laser beam is imaged onto the pupil of an objective lens (Mitutoyo G Plan Apo 50$\times$) via a dichroic mirror reflection, then focused onto the atoms, and imaged by another the same objective lens for monitoring. At the atoms, the tweezer beam has 0.9 $\rm\mu m$ waist (measured by trap frequency) and linear polarization of vertical direction. The dichroic mirror transmits 556 nm fluorescence from atoms collected by the objective lens, and the fluorescence is imaged onto an EMCCD camera (Andor iXon Ultra 897) with 50$\times$ magnification. The total collection efficiency of the fluorescence into $5\times5$ pixels ROI in the camera is 2.24\%, measured by comparing estimated scattered photons and detected photons.

To load atoms, we firstly turn on the tweezers of 0.2 mK trap depth or higher, to the compressed MOT. As turning off the MOT, we ramp down the tweezers to 0.1 mK trap depth and apply green MOT beams with power 2 mW per axis and detuning +3 MHz from ${^1S_0},F=1/2-{^3P_1},F=3/2,m_F=+1/2$ resonance for 200 ms in B field of 15 G and a tilt angle of about $15.5^\circ$ (1st-order magic angle) from vertical axis (y axis). This drives blue-detuned light assisted collisions (LAC) resulting in about 80\% single atom loading rate. We have two cooling 556 nm beams controlled by their own double-pass AOMs driven by a multi-tone signal generator (ARTIQ Sinara 4624 Phaser). One beam is applied to the atoms with a tilt angle $52^\circ$ from the tweezer axis (z axis) in horizontal plane (xz plane), and the other beam is irradiated by a tilt angle $73^\circ$ in vertical plane ($\sim$yz plane), and both beams are retro-reflected, respectively. These cooling beams are used for red-detuned LAC in 0.2 mK trap depth for 100 ms to make sure of eliminating multi-atom loading, and imaging of single atoms as described in the main text. 1389 nm repumping laser (Toptica DL Pro) for ${^3P_0}-{^3D_1}$ transition is frequency-locked to a ULE cavity, and applied to the atoms with all polarization components and intensity 28 $\rm mW/cm^2$. 770 nm repumping laser for ${^3P_2}-{^3S_1}$ transition is a homemade ECDL frequency-locked to the wavemeter. The beam is modulated by a fiber EOM to generate sidebands for simultaneously addressing $F=5/2$ and $F=3/2$ states, and illuminates the atoms with all polarization components and intensity 45 $\rm mW/cm^2$.

In order to generate SLM phase for atom array, we use weighted Gerchberg-Saxton algorithm and feedback for homogenization~\cite{Kim2019}. $3\times3$ square array of 5$~\rm\mu m$ spacing is generated and homogenized based on the monitor camera, at first. Then, we do spectroscopy of ${^1S_0},F=1/2-{^3P_1},F=3/2,m_F=+3/2$ transition to measure the in-situ trap depth, and do the feedback several times, resulting in uniform trap depths within 2\% standard deviation.

\section{Morris-Shore transformation for $\bf ^1S_0,F=1/2~-~^3P_1,F=3/2,\abs{m_F}=1/2$ transition}
\label{sec_MS}
When both $\ket{\pm1/2'}\equiv\ket{^3P_1,F=3/2,m_F=\pm1/2}$ are simultaneously driven from $\ket{\pm1/2}\equiv\ket{^1S_0,F=1/2,m_F=\pm1/2}$ by dual-tone laser beams of the same intensity and detuning $\Delta$ for each transition, the system of interest is a double lambda system of $\ket{\pm1/2}$ and $\ket{\pm1/2'}$ of which Hamiltonian is given as
\begin{align}
\hat H = \begin{pmatrix}
0 & 0 & \Omega_{\pi}^* & \Omega_{\sigma_+}^*\\
0 & 0 & \Omega_{\sigma_-}^* & \Omega_{\pi}^*\\
\Omega_{\pi} & \Omega_{\sigma_-} & \Delta & 0\\
\Omega_{\sigma_+} & \Omega_{\pi} & 0 & \Delta
\end{pmatrix},
\end{align}
where $\Omega_i$ is Rabi frequency for i polarization component and energy splitting between $\ket{\pm1/2}$ is neglected. Morris-Shore (MS) transformation~\cite{MS1983} allows to represent this system as independent two-state systems, $\ket{g_-}-\ket{e_-}$ and $\ket{g_+}-\ket{e_+}$, by diagonalization of off-diagonal blocks, $V^\dagger=\begin{pmatrix}\Omega_{\pi} & \Omega_{\sigma_-}\\
\Omega_{\sigma_+} & \Omega_{\pi}\end{pmatrix}$ and $V$. It is known that the Rabi frequency $\Omega_\pm$ for the MS-transformed system $\ket{g_\pm}-\ket{e_\pm}$ is given by square root of eigenvalues of $VV^\dagger$ or $V^\dagger V$, expressed as
\begin{widetext}
\begin{align}
\Omega_{\pm}=\sqrt{{\abs{\Omega_{\pi}}^2+\frac{1}{2}\abs{\Omega_{\sigma_+}}^2+\frac{1}{2}\abs{\Omega_{\sigma_-}}^2}\pm \sqrt{\frac{1}{4}(\abs{\Omega_{\sigma_+}}^2-\abs{\Omega_{\sigma_-}}^2)^2+\abs{\Omega_{\pi}}^2(\abs{\Omega_{\sigma_+}}^2+\abs{\Omega_{\sigma_-}}^2)+2{\rm Re}(\Omega_{\pi}^2\Omega_{\sigma_+}^*\Omega_{\sigma_-}^*)}}.
\label{eq_MS}
\end{align}
\end{widetext}
Note that transformation matrix from $\ket{\pm1/2}$ ($\ket{\pm1/2'}$) to $\ket{g_\pm}$ ($\ket{e_\pm}$) can be obtained by transformation matrix to diagonalize $VV^\dagger$ ($V^\dagger V$). For the case where the k vector of driving laser beam and the quantization axis are perpendicular, $\Omega_{\sigma_+}=-\Omega_{\sigma_-}$ can hold true by setting x axis as the direction of E field component perpendicular to the quantization axis, and Eq.~\ref{eq_MS} is reduced to Eq.~\ref{eq_MSRabi}.

\section{Steady-state solution for dual-tone narrowline cooling}
\label{sec_steady}
For the special case $\phi=0,\pi/2$ in the Sec.~\ref{sec_dualtone}, Hamiltonian can be written in real numbers and there is symmetry between $m_F=\pm1/2$. Then, optical Bloch equations~\cite{GAFbook} in the MS-transformed basis (see Sec.~\ref{sec_MS}) are given as
\begin{align}
&\frac{d\rho_{g_-g_-}}{dt}=f\Gamma\rho_{e_-e_-}+(1-f)\Gamma\rho_{e_+e_+}+\frac{i\Omega_-}{2}(\rho_{e_-g_-}-\rho_{g_-e_-})\nonumber\\
&\frac{d\rho_{e_-e_-}}{dt}=-\Gamma\rho_{e_-e_-}+\frac{i\Omega_-}{2}(\rho_{g_-e_-}-\rho_{e_-g_-})\nonumber\\
&\frac{d\rho_{g_-e_-}}{dt}=-(\Gamma/2+i\Delta)\rho_{g_-e_-}+\frac{i\Omega_-}{2}(\rho_{e_-e_-}-\rho_{g_-g_-})\nonumber\\
&\frac{d\rho_{g_+g_+}}{dt}=(1-f)\Gamma\rho_{e_-e_-}+f\Gamma\rho_{e_+e_+}+\frac{i\Omega_+}{2}(\rho_{e_+g_+}-\rho_{g_+e_+})\nonumber\\
&\frac{d\rho_{e_+e_+}}{dt}=-\Gamma\rho_{e_+e_+}+\frac{i\Omega_+}{2}(\rho_{g_+e_+}-\rho_{e_+g_+})\nonumber\\
&\frac{d\rho_{g_+e_+}}{dt}=-(\Gamma/2+i\Delta)\rho_{g_+e_+}+\frac{i\Omega_+}{2}(\rho_{e_+e_+}-\rho_{g_+g_+}),
\label{eq_OBE}
\end{align}
where $f\in[0,1]$ is a fixed number determined by laser polarization and $\Gamma=2\pi\times182$ kHz is the natural linewidth of the excited states. For a steady-state solution, the time derivatives in Eq.~\ref{eq_OBE} are set to 0, resulting in
\begin{align}
&\rho_{e_-e_-}-\rho_{e_+e_+}=0\nonumber\\
&\frac{1+s_-+(2\Delta/\Gamma)^2}{s_-}\rho_{e_-e_-}+\frac{1+s_++(2\Delta/\Gamma)^2}{s_+}\rho_{e_+e_+}=1,
\end{align}
where $s_\pm=2\Omega_\pm^2/\Gamma^2$ is intensity saturation parameter. Using these relations, total scattering rate $R_{\rm tot}=\Gamma(\rho_{e_-e_-}+\rho_{e_+e_+})$ leads to Eq.~\ref{eq_scattering}. This result can be interpreted as following. The ratio between the populations of two ground states $\ket{g_\pm}$ is determined by the ratio between the inverses of pumping out rates from the states, $1/R_\pm$, where $R_\pm=\frac{\Gamma}{2}\frac{s_\pm}{1+s_\pm+(2\Delta/\Gamma)^2}$ is two-level scattering rate for $\ket{g_\pm}-\ket{e_\pm}$ transition. Then, the total scattering rate is given by $R_{\rm tot}=\frac{R_-}{R_+ + R_-}\times R_+ + \frac{R_+}{R_++R_-}\times R_-=\frac{2R_+R_-}{R_+ + R_-}$.

\section{Monte Carlo simulation of laser cooling dynamics in tweezer}
\label{sec_MC}

\begin{figure}
\includegraphics[width=0.48\textwidth]{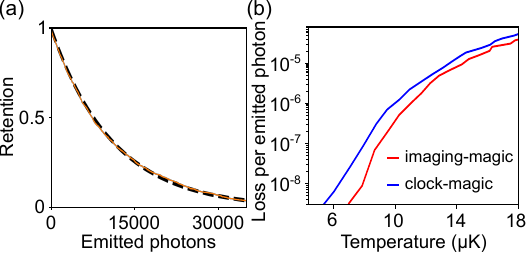}
\caption{(a) Time dependence of atom retention during 2-axis cooling from MC simulation with $s=2$, $\Delta=-\Gamma$, in clock-magic tweezer of $0.1~{\rm mK}$ trap depth. Black dahsed line is an exponential fitting for the simulation data. (b) Loss per emitted photon during 2-axis cooling depending on temperature, calculated by MC simulation as varying s with $\Delta=-\sqrt{1+s}~\Gamma/2$ in imaging-magic tweezer of $0.2~{\rm mK}$ trap depth, and with $\Delta=-\sqrt{1+s}~1.15\Gamma/2$ in clock-magic tweezer of the same depth. In (a) and (b), imaging polar angles for two axes are $\theta_k=52^\circ,73^\circ$, respectively.}
\label{fig_MC}
\end{figure}

In this section, we describe Monte Carlo (MC) simulation of atom's laser cooling dynamics in tweezer. The simulation treats atom's motion classically, based on forces from a tweezer trap of Gaussian beam potential,
\begin{align}
U(\vec{r})=-\frac{U_0}{1+(z/z_R)^2}\exp\left(-\frac{2(x^2+y^2)}{w_0^2(1+(z/z_R)^2)}\right),
\end{align}
with trap depth $U_0$, beam waist $w_0$, and Rayleigh range $z_R=\pi w_0^2/\lambda_{\rm trap}$, and forces from stead-state photon scattering rate as in conventional Doppler cooling theory~\cite{lasercooling} where interference between cooling beams is not considered.

At first, atom's position and velocity are set randomly by Maxwell-Boltzmann distribution of the initial temperature. At each time, detuing is updated by position-dependent differential AC Stark shift for general non-magic traps, as $\Delta \rightarrow \Delta'=\Delta+\gamma(U_0+U(\vec{r}))/\hbar$, where $\gamma$ is relative differential AC Stark shift of the excited state, which results in photon scattering rate for a pair of counter-propagating laser beams along i-th axis as
\begin{align}
R_i = \frac{\Gamma}{2}\frac{s_i}{1+\sum_i s_i+(2\Delta'/\Gamma)^2}.
\end{align}
Then, velocity is updated by $\vec{v}\rightarrow \vec{v}+\left(-\vec{\nabla} U'(\vec{r})-\alpha\vec{v}\right)\delta t/m+\hat{e}_p\sqrt{2D_p \delta t}/m$, where effective trap depth is $U'(\vec{r})=(1-\sum_i R_i/\Gamma)U(\vec{r})+(\sum_i R_i/\Gamma)(1+\gamma)U(\vec{r})$, damping coefficient is
\begin{align}
\alpha = -4\hbar k^2\frac{\sum_i s_i(\vec{k_i}\cdot\vec{v}/k_iv)^2}{2}\frac{2\Delta'/\Gamma}{\left[1+\sum_i s_i+(2\Delta'/\Gamma)^2\right]^2}
\end{align}
with k vector for i-th axis $\vec{k_i}$, and diffusion constant is
\begin{align}
D_p=\hbar^2k^2\sum_i R_i
\end{align}
with randomly chosen unit vector $\hat{e}_p$ considering dipole radiation patterns with respect to the quantization axis (y axis). After updating velocity, position is updated by $\vec{r}=\vec{r}+\vec{v}\delta t$. To obtain atom retention, we regard an atom as lost if its total energy is not negative. We repeat this MC simulation with time increment $\delta t=1~{\rm\mu s}$ and get average values to compare them with experiments.

\begin{figure}
\includegraphics[width=0.48\textwidth]{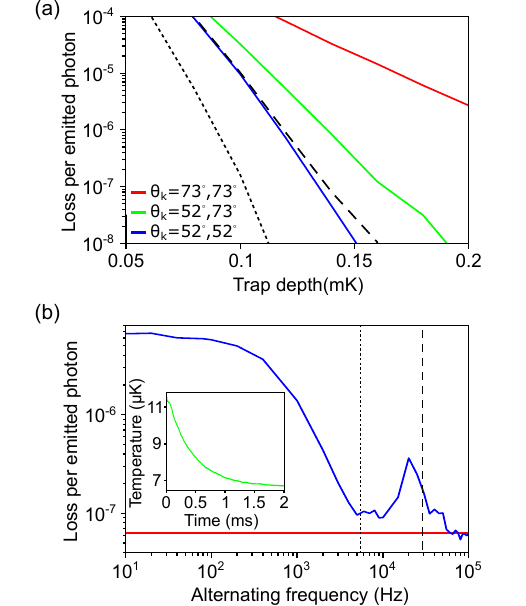}
\caption{(a) Loss per emitted photon for 2-axis cooling as a function of the depth of imaging-magic trap, with $s=1,\Delta=-\sqrt{2}\Gamma/2$, and various $\theta_k$, in cooling-magic tweezer. Dashed line is a result with isotropic 3-axis cooling of the same $s$ and $\Delta$, and dotted line indicates a 2-axis result with $s=0.1,~\Delta=-\Gamma/2,~\theta_k=52^\circ,52^\circ$, which could be regarded as a lower limit. (b) Loss dependence on alternating frequency for 2-axis alternating cooling in 0.14 mK imaging-magic trap with $s=1,~\Delta=-\sqrt{2}\Gamma/2,~\theta_k=52^\circ,52^\circ$. Blue solid line is the simulation result and red solid line indicates the result from 2-axis simultaneous cooling. Dashed (dotted) line corresponds to radial (axial) trap frequency. Inset shows atom temperature dynamics of 1-axis cooling with the same parameter, from an initial temperature.}
\label{fig_MC2}
\end{figure}

In Fig.~\ref{fig_MC}, we observe some characteristics of atom loss during the laser cooling dynamics utilizing the MC simulation. First, as shown in Fig.~\ref{fig_MC}(a), the atom loss is well fitted by an exponential fitting, which means the recoil loss rate is constant during the dynamics. This is different from the non-exponential initial behavior of the loss induced by increased temperature from heating~~\cite{heating}. This is why we use loss per photon as a metric to measure the loss from momentum damping and diffusion dynamics in the main text. Second, the atom loss has strong positive correlation with temperature as plotted in Fig.~\ref{fig_MC}(b).

Now, we search more optimal configuration for 2-axis cooling using the MC simulation. The atom loss shown in Fig.~\ref{fig_MC2}(a) is severely affected by imaging polar angle $\theta_k$. As the optimal temperature for 2-axis cooling is around $52^\circ$ in Fig.~\ref{fig_cooling}(a), $\theta_k=52^\circ,52^\circ$ exhibits the best performance. Remarkably, this optimal 2-axis cooling shows slightly better performance than that of isotropic 3-axis cooling, probably due to the anisotropy from tweezer's looser axial direction, and from the spontaneous emission pattern. With $\theta_k=52^\circ,52^\circ$, trap depth can be reduced to about 0.14 mK, where the summation of the losses from the MC simulation and trap Raman scattering (Appendix~\ref{sec_scattering}) is minimized to about $1\times10^{-6}/$photon. For $s=0.1$ and $\Delta=-\Gamma/2$, it can be further reduced to about 0.11 mK at the expense of more than 5 times longer imaging time. Finally, in Fig.~\ref{fig_MC2}(b), performance of 2-axis alternating cooling in 0.14 mK depth is compared with 2-axis (simultaneous) cooling, depending on alternating frequency. As shown in the inset, thermal equilibration time of 1-axis cooling is $\sim 1~{\rm ms}$, and alternating frequency above $\sim 1~{\rm kHz}$ exhibits performance comparable to the simultaneous cooling. Note that there is a higher loss region near the radial trap frequency resonance, which should be avoided for the optimal performance.

\section{Calculation of trap Raman scattering to metastable states}
\label{sec_scattering}
At high trap depth, Raman scattering from $^3P_1,F=3/2,\abs{m_F}=1/2$ to metastable states $^3P_0$ and $^3P_2$ by 759.356 nm trap photon is the dominant loss source for narrowline imaging~\cite{Kaufman2023}. Let the former and latter Raman scattering rates be $\Gamma_{1\rightarrow0}$ and $\Gamma_{1\rightarrow2}$, respectively. Then, the loss rate from Raman scattering during imaging is given by
\begin{align}
\Gamma_{\rm Raman}=&~~(\Gamma_{1\rightarrow0}+\Gamma_{1\rightarrow2})(\rho_{e_-e_-}+\rho_{e_+e_+})\nonumber\\
\equiv&~~L_{\rm Raman}R_{\rm tot},
\end{align}
where $L_{\rm Raman}=(\Gamma_{1\rightarrow0}+\Gamma_{1\rightarrow2})/\Gamma$ is Raman scattering loss per emitted imaging photon. Raman scattering rates~\cite{Lisdat2018} by trap laser, $E_l\cos\omega_l t$, can be calculated as
\begin{widetext}
\begin{align}
\Gamma_{1\rightarrow f}=\frac{E_l^2\omega_{\rm sc}^3\Theta(\omega_{\rm sc})}{12\pi\epsilon_0\hbar^3 c^3}\sum_{m_F=\pm\frac{1}{2}}\sum_{F',m_{F'}}\sum_q\Bigg|\frac{1}{2}\sum_k\frac{\bra{^3P_f,F',m_{F'}}d_q\ket{k}\bra{k}d_l\ket{^3P_1,F=3/2,m_F}}{\omega_{1k}-\omega_l}\nonumber\\
+\frac{\bra{^3P_f,F',m_{F'}}d_l\ket{k}\bra{k}d_q\ket{^3P_1,F=3/2,m_F}}{\omega_{1k}+\omega_{\rm sc}}\Bigg|^2,
\end{align}
\end{widetext}
where $\Theta(x)$ is Heaviside step function, $\omega_{\rm sc}=\omega_l-\omega_{1f}$ is the frequency of scattered photon given by trap laser frequency $\omega_l$ and the transition frequency between $\ket{i}$ and $\ket{j}$, $\omega_{ij}$, $d_l$ and $d_q$ are components of electric dipole operator along polarizations of trap laser and spontaneous emission, respectively. Note that trap Raman scattering rate is proportional to trap depth, i.e. $\Gamma_{\rm Raman}\propto E_l^2\propto U_0$. Using known dipole matrix elements, calculated value of $L_{\rm Raman}/U_0$ is about $6.6\times10^{-6}~{\rm photon^{-1}\cdot mK^{-1}}$ and this is compared with experiments in the main text.
\end{appendix}

\end{document}